\documentclass[extra,onecolumn,doublespacing]{gji}
\usepackage{timet,color,graphicx}
\usepackage{lineno}
\usepackage{graphicx}
\usepackage{float}
\usepackage[final]{hyperref}
\hypersetup{
    colorlinks = true,
    linkcolor = blue,
    anchorcolor = blue,
    citecolor = blue,
    filecolor = blue,
    urlcolor = blue
}

\title[Cloud Seismology]
  {A Review of Cloud Computing in Seismology}

  \author[Y. Ni et al.]
  {Yiyu Ni$^1$, Marine A. Denolle$^1$, Jannes Münchmeyer$^2$, Yinzhi Wang$^3$, Kuan-Fu Feng$^{1,4}$, \and Carlos Garcia Jurado Suarez$^5$, Amanda M. Thomas$^6$, Chad Trabant$^7$, \and Alex Hamilton$^7$, and David Mencin$^7$ \\
    $^1$ Department of Earth and Space Sciences, University of Washington, 4000 15th Ave NE, Seattle, WA 98195 USA \\
    $^2$ Univ. Grenoble Alpes, Univ. Savoie Mont Blanc, CNRS, IRD, Univ. Gustave Eiffel, ISTerre, CS 40700 38058 GRENOBLE Cedex 9, France\\
    $^3$ Texas Advanced Computing Center, University of Texas, 10100 Burnet Rd, Austin, TX 78758, USA \\
    $^4$ Department of Geology and Geophysics, University of Utah, 115 S 1460 E, Salt Lake City, UT 84112, USA \\
    $^5$ eScience Institute, University of Washington, Campus Box 351570, 3910 15th Ave NE, Seattle, WA 98195, USA \\
    $^6$ Department of Earth and Planetary Sciences, University of California, One Shields Avenue, Davis, California 95616, USA \\
    $^7$ EarthScope Consortium, 1200 New York Avenue NW, Suite 400, Washington DC, 20005, USA \\
    }
    
\date{\today}

\pagerange{\pageref{firstpage}--\pageref{lastpage}}
\volume{X}
\pubyear{{\the\year{}}}

\begin{document}
\label{firstpage}

\maketitle

\begin{summary}

Seismology has entered the petabyte era, driven by decades of continuous recordings of broadband networks, the increase in nodal seismic experiments, and the recent emergence of Distributed Acoustic Sensing (DAS). This review explains how commercial clouds — AWS, Google Cloud, and Azure — by providing object storage, elastic compute, and managed databases, enable researchers to “bring the code to the data,” thereby overcoming traditional HPC solutions' bandwidth and capacity limitations. After literature reviews of cloud concepts and their research applications in seismology, we illustrate the capacities of cloud-native workflows using two canonical end-to-end demonstrations: 1) ambient noise seismology and cross-correlation, and 2) earthquake detection, discrimination, and phase picking. Both workflows utilized S3 for streaming I/O and DocumentDB for provenance, demonstrating that cloud throughput can rival on-premises HPC at comparable costs, scanning 100 TBs to 1.3 PBs of seismic data in a few hours or days of processing. The review also discusses research and education initiatives, the reproducibility benefits of containers, and cost pitfalls (e.g., egress, I/O fees) of energy-intensive seismological research computing. While designing cloud pipelines remains non-trivial, partnerships with research software engineers enable converting domain code into scalable, automated, and environmentally conscious solutions for next-generation seismology. 

\end{summary}


\section{Introduction}
Seismology has entered a “petabyte era,” where seismic networks and nodal array experiments routinely generate more data than traditional workstations and institutional clusters can store or analyze. Data from more than 70,000 seismometers has surpassed 1 PB on the EarthScope Data Archive \citep{arrowsmith2022big}. Novel technologies, such as Distributed Acoustic Sensing (DAS), which collects thousands of sensors per experiment, are also increasing the data storage needs and already surpassing the PBs of data (i.e., datasets exceeding $10^15$ bytes) collected and shared \citep{zhan2020distributed, spica2023pubdas, wuestefeld2024global}. Moving large amounts of data is challenging in part due to hardware limitations of disk I/O and in part due to throughput speeds being constrained by internet bandwidth. Moreover, the current well-adopted seismic data formats and services are not well suited for big-data seismology studies \citep{quinteros2021exploring, arrowsmith2022big}. While seismologists' workflows traditionally involve analyzing data by downloading from archives and working on-premise, this model faces significant difficulties with analysis that requires more than several TBs of data. Researchers are exploring cloud computing to address these bottlenecks, which brings code to data and offers scalable storage and processing. 

To turn this data deluge into discovery, seismologists increasingly look to cloud platforms that put compute next to the archive. {\it Cloud computing} refers to using remote data centers to store data and run computations on demand via the internet. Commercialized cloud computing emerged in the early 2000s, when the industry began providing capabilities such as large-scale object storage and on-demand computing (e.g., Amazon Web Services (AWS), Google Cloud Platform (GCP), Microsoft Azure). Geophysicists have traditionally used high-performance computing (HPC) centers to deliver tightly coupled, job-scheduled computing on a shared filesystem, often driven by large-scale numerical simulations. In contrast to HPC systems, cloud providers deliver elastic resources as metered services that users spin up and pay for only when needed. This encompasses several key design elements, including virtual machines (VMs), cloud storage, and advanced services like databases and serverless services. Harnessing cloud infrastructure could fundamentally change how seismologists handle big data, making analyses faster and more collaborative. 

\begin{figure}
    \centering
    \includegraphics[width=0.75\linewidth]{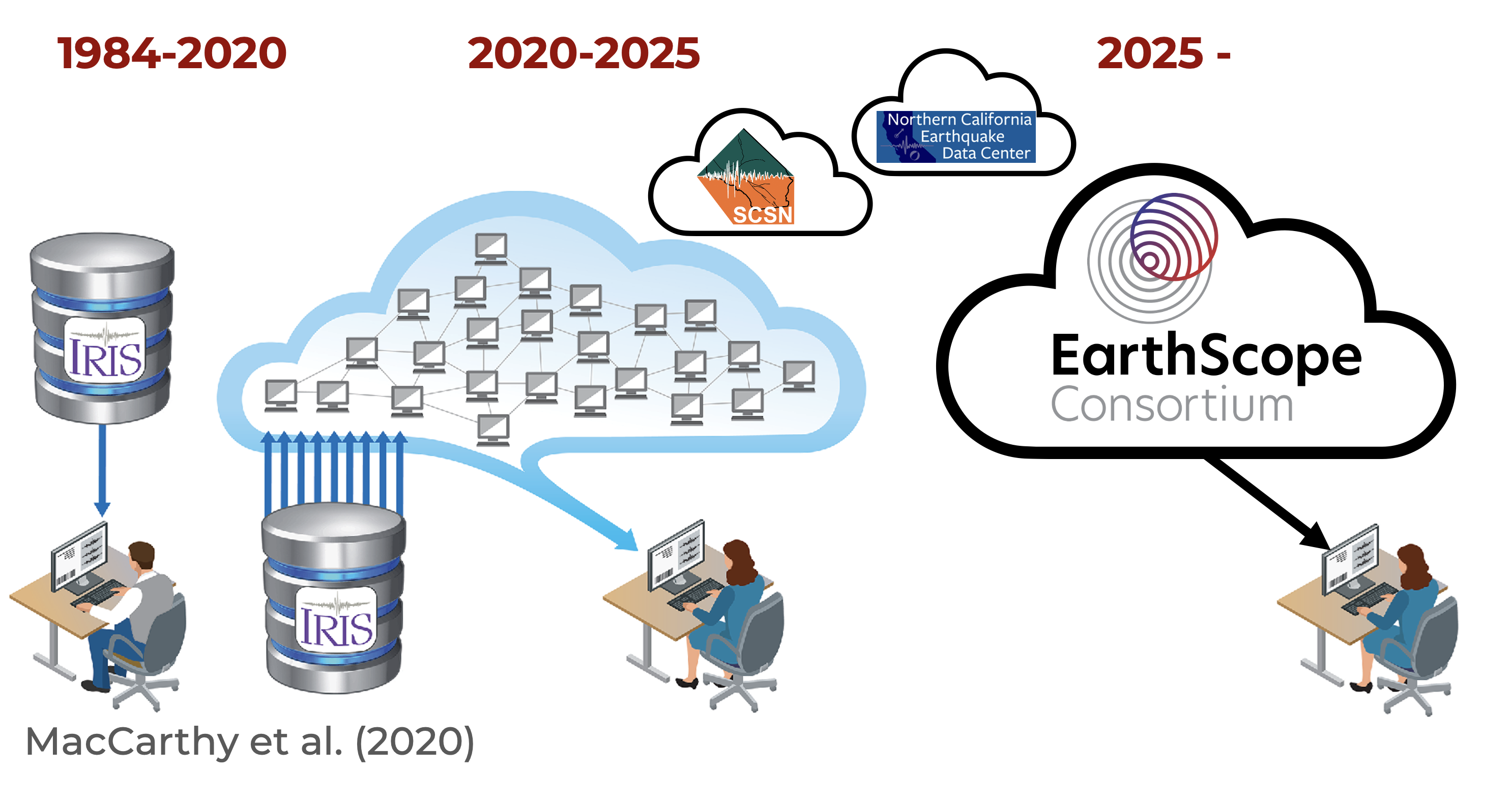}
    \caption{Evolution of cloud computing in seismology: before 2020, most computational workflows involved downloading data from the Incorporated Research Institutions for Seismology Data Management Center (IRIS DMC) and other data centers and working locally. Between 2020 and 2025, seismologists have investigated the use of elastic computing by pulling data from existing archives and processing directly on the cloud \citep[e.g.,][]{maccarthy2020seismology}. At the same time, two regional seismic networks copied their archives of earthquake catalogs and seismic waveforms to Amazon Web Services (AWS): the Southern California Seismic Network (SCSN) and the Northern California Seismic Network (NCSN). Since 2025, the EarthScope Consortium has migrated its petabyte-scale archive to the cloud, enabling researchers to pull and compute directly on the cloud.}
    \label{fig:history}
\end{figure}

Cloud offers tremendous opportunities for easy access to object storage, a centralized, affordable, and widely accessible solution for massive data archives. In contrast to HPC, object storage enables hosting and sharing PBs of public geoscientific data \citep{zhuang2020enabling, abernathey2021cloud, gentemann2021science}. While individual data queries may have modest throughput speeds (10-1000 MB/s), the immense parallelization capabilities allow throughput speeds comparable to those of HPC scratch systems (10-100 GB/s). Moreover, cloud providers are storing PBs of publicly available and free-access geoscientific data \citep{abernathey2021cloud}. In seismology, cloud-hosted seismic data is rising as we illustrate in Figure~\ref{fig:history}. The Southern California Earthquake Data Center (SCEDC) was the first to provide archives of regional seismic networks as open datasets on the commercial cloud \citep{yu2021scedc}. It was recently followed by the Northern California Earthquake Data Center (NCEDC) and the EarthScope-operated Seismological Facility for the Advancement of Geoscience repository (formerly Incorporated Research Institutions for Seismology Data Management Center, IRIS DMC). Cloud has also been a promising storage solution for DAS data (e.g., PoroTomo, \url{https://registry.opendata.aws/nrel-pds-porotomo/}, and Ridgecrest DAS \citep{yu2021scedc}). 

Open-source notebooks (e.g., Binder, Google Colab, and EarthScope GeoLab) lower the entry barrier to the cloud by giving users a ready-made Python environment next to the data. For example, a typical entry point for scientists to cloud computing is through a freely accessible JupyterHub with a backend running on cloud platforms. Users can automate software-to-infrastructure, using tools such as {\tt repos2docker} to containerize software and automatically give access to cloud-hosted virtual machines. In particular, \citet{krischer2018seismo} has pioneered the use of cloud-hosted Jupyter notebooks, utilizing modest yet free Binder instances provided and donated by various cloud providers (e.g., OVHcloud as of 2025). Alternatively, Google Colab provides free access to modest-sized virtual machines, which utilize GCP resources with pre-defined Python environments. Cloud is also an on-demand platform to host educational materials \citep{denolle2024training}. 

Considering these basics, this article reviews how cloud computing has been applied in seismological research. We discuss storage solutions fro seismic data (Section~\ref{sec:storage}), databases (Section~\ref{sec:db}), the various types of computing resources (Section~\ref{sec:compute}), showcase experiments conducted at scale on large archives of broadband seismic data (Section~\ref{sec:ex}), and cloud-enabled visualizations (Section~\ref{sec:viz}). We also present a series of experiments that have targeted archetypes in big-data seismology: 1) data mining using deep learning models to detect seismic events, and 2) ambient field seismology that requires intensive generation of cross-correlation at scale. Both tasks are characterized by a high data intake and large computational requirements, yet differ in output and processing specifics. In particular, we focus on the specific requirements of each workflow and how they affect the choice of cloud tools. Finally, we present our recent experience in running a cloud seismology workshop (Section~\ref{sec:edu}) and discuss the opportunities and challenges in this area (Section~\ref{sec:discussion}).

\section{Cloud Storage}\label{sec:storage}
\subsection{Cloud-hosted Data Archives}
Transferring and sharing data between institutions is essential for large-scale collaborative research in seismology. Software from command-line tools (e.g., {\tt sftp}, {\tt wget}) to hosted services (e.g., Globus, \citet{allen2012software}) is made available to securely and efficiently share data over the internet to facilitate the transfer of moderate to large-sized data sets. However, on-premise storage and bandwidth limitations constrain point-to-point transfer efficiency. At the same time, the host takes responsibility for processing requests and maintaining stable data access, usually at the cost of additional man-power. Cloud storage provides an accessible solution for hosting and sharing scientific data with exceptional scalability and durability, along with improved findability, accessibility, and reusability \citep{abernathey2021cloud}. 

Cloud object storage has the significant advantage of storing large datasets while being scalable to massively parallel queries from within and outside the cloud. For instance, AWS provides the Simple Storage Service (S3), an object storage enabling users to store virtually unlimited amounts of data in any format. Azure (Blob storage) and GCP also provide similar services. In contrast to the filesystem, files are saved as individual objects and organized in a bucket with a flat structure. S3 objects could be located with a prefix and a key, an analogy to the folder and the filename that users are more familiar with in a filesystem. While S3 is not a Portable Operating System Interface and differs from the Unix-style semantics most researchers are used to, the departure is advantageous: freeing storage from strict POSIX rules lets cloud object stores scale to billions of files, offer global access from any service, and replicate transparently across regions.

Seismic data centers curate and offer vast amounts of invaluable data through the International Federation of Digital Seismograph Stations (FDSN) web services, which standardized API queries for seismic data and metadata delivery \citep{hutko2017data, quinteros2021geofon, hauksson2020caltech}. Cloud storage fundamentally transforms data center solutions, improving robustness, durability, and stability to data management within the facility, resilience to storage read spikes, and proximity between storage and compute nodes on the user end \citep{beckwith2011managing}. The SCEDC has been the pioneer in migrating, hosting, and publicizing $\sim$150 TB of continuous data on AWS S3 \citep{yu2021scedc, zhu2025california}, followed by the effort of NCEDC of their $\sim$190 TB, and most recently, EarthScope's seismological data archive, which has surpassed 1 PB. Direct access to cloud-based archives enables the development of cloud-native workflows, which we will discuss in Section~\ref{sec:ex}. For instance, the workflow's throughput could not be achieved through the FDSN \texttt{fdsnws-dataselect} web service.

Data centers migrating their repositories, or offering copies, in cloud environments must consider how data will be discovered and accessed directly by researchers. Whereas previously, these repositories were only accessible behind services that acted as abstraction layers, the organizations' data is now exposed to direct access. This is important to avoid adding processing bottlenecks and limit the ability to subset data. Operational requirements like controlling access to restricted data, log data use, etc., are now bottlenecks and should be minimized. New services and software are needed to support the efficient discovery and use of large-scale analyses beyond the simplest cases.

\begin{figure}
    \centering
    \includegraphics[width=\linewidth]{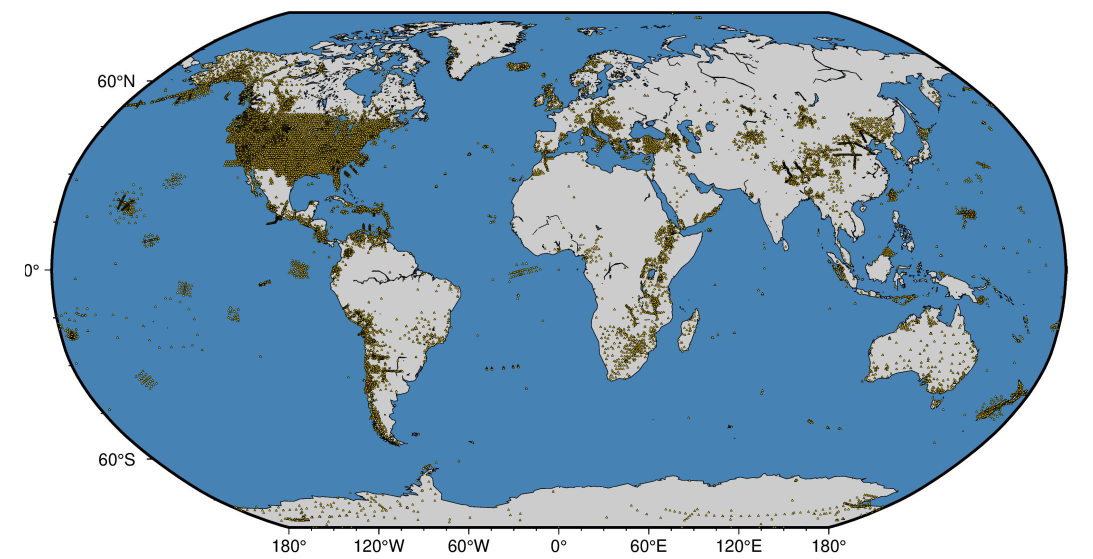}
    \caption{{\bf Station map of cloud-hosted data}: A total of about 1.3 PB of seismometer miniSEED data is hosted on AWS Simple Storage Service (S3): the EarthScope Consortium archive, and the Northern and Southern California data. Each triangle indicates a seismic station.}
    \label{fig:data}
\end{figure}

\subsection{Data Formats on Object Storage}
\subsubsection{Array Data}
Historically, data centers have relied on SEG-Y (for active seismology) and Standard for the Exchange of Earthquake Data (SEED, for passive seismology) as primary data formats \citep{guimaraes2021high}. SEG-Y, designed initially for tape storage in the 1970s, remains widely used due to regulatory requirements in the petroleum exploration industry. Still, it suffers from poor parallel read performance and high I/O latency in modern cloud environments. Meanwhile, SEED (or more accurately miniSEED) is also actively used for seismic time series archiving and shipping. Specifically, the miniSEED standard in its 2+ version adopted the paradigm that separates waveforms (i.e., miniSEED with minimal metadata) from their metadata counterpart (i.e., dataless SEED with no time series). This only makes this format partially cloud optimized because it has no native support for object-based data partitioning, indexing, or scalable metadata integration. It also requires additional infrastructure to serve efficiently from cloud storage (e.g., indexing layer or metadata catalog).

More recently, the HDF5 format \citep[Hierarchical Data Format version 5,][]{hdf5cite} enables researchers to design their data structures and storage. HDF5 supports a customizable and flexible hierarchical schema, allowing the storage of multidimensional seismic waveforms and metadata \citep[e.g., ][]{krischer2016adaptable, white2023hdf5eis}. On the other hand, such a self-describing structure also presents obvious limitations when used at scale in distributed cloud computing. The monolithic nature of HDF5 files introduces overhead for metadata handling and parallel access, while losing efficiency on byte-range requests and compressed blocks, i.e., read subsets of large datasets \citep{ni2023object}. Despite efforts from open-source projects such as H5coro (H5 cloud-optimized read-only library, \url{https://github.com/SlideRuleEarth/h5coro}) and kerchunk (\url{https://github.com/fsspec/kerchunk}) that were made to optimize HDF5 for cloud object storage, they are often read-only solutions and do not fully resolve challenges from generalized file structures. 

Recognizing these limitations, geospatial data initiatives such as Pangeo and EarthCube have pioneered the adoption of cloud-optimized array formats like Zarr and TileDB, which avoid hierarchical formats and instead store multi-dimensional arrays in a chunked, compressed, and distributed manner. Pangeo's use of Zarr, for instance, has enabled massive parallel processing of gridded climate and remote sensing datasets, demonstrating $\sim$10x speedup in read performance compared to HDF5 when accessed in parallel from cloud object stores \citep{abernathey2021cloud}. Similarly, TileDB has proven effective for sparse geospatial data, including GNSS and seismic sensor arrays, allowing efficient subsetting and time-series access \citep{habermann2021common}. These technologies are now being adapted for seismology, where efforts such as those by \citet{ni2023object} demonstrate that converting DAS data from HDF5 to Zarr/TileDB results in significant memory and compute time improvements. The proliferation of these new open formats and their variants may be an obstacle to the sustainability of our software, so seismological workflows must stay format-agnostic and ready to pivot between, or simultaneously support, multiple storage layouts as standards evolve.

\subsubsection{Point Sensors}
Seismological research spans a wide range of temporal and spatial scales, requiring storage solutions that can support both short-duration, multi-station queries and long-duration, station-centric analyses. These workflows stress storage along different query axes: some pull data from thousands of stations but only short time windows (e.g., large-N arrays, ambient-noise cross-correlation), whereas others retrieve long, continuous histories—months to years—for each station in the network (e.g., ambient-noise monitoring, template matching). Small-object storage enables efficient writing and retrieval, often providing superior performance relevant to seismological broadband seismic data (e.g., 10 MB per day for a typical 100 Hz 3-component broadband data). 

As an aside, there are formats designed for petabyte+ data sets, cloud-optimized formats like Zarr and TileDB that are designed for large queries on colocated compute with efficient streaming capabilities for large data chunks, convenient for geospatial data and array-based data such as nodal arrays and distributed acoustic sensing data \citep[e.g.,][]{ni2023object}. In contrast, the older formats, e.g., the day-long miniSEED files, permit efficient remote clients accessing small data slices and may be ideal for real-time data streaming of single-station-based data access and event-driven workflows. In principle, they allow for random access into the waveforms at blockette-granularity, though no cloud-native access library exists to this date. 

The data centers mentioned above all host seismic time series in the miniSEED format on AWS S3. There are differences in conventions for organizing files in the S3 bucket regarding data granularity. Specifically, SCEDC and NCEDC store one channel per object, whereas the EarthScope archive groups all channels per station in a single file. Despite hosting more small objects, the former structure exhibits high efficiency when querying data from subset channels or locations since no redundant bytes are read. The latter structure usually requires an external database that indexes files to facilitate data query (e.g., using mseedindex, \url{https://github.com/EarthScope/mseedindex}). The various bucket structures indicate that object naming (prefix and key) has not yet reached a standard across these data centers because the new paradigm of direct access renders these previously hidden implementation details part of the user interface. 

The metadata counterpart has not been well adapted to the cloud. The FDSN web service for the metadata still runs on on-prem servers at each data center, which is not resilient to scaled queries. The first step from SCEDC and NCEDC is to provide community-standardized StationXML files in their respective S3 bucket containing the full history of station metadata and instrumental response. Reading and parsing full StationXML files through S3 is considered less efficient than querying through the FDSN web service. There remains space for improvement in hosting and delivering metadata efficiently. 

\section{Cloud Databases}\label{sec:db}
Seismologists need databases to manage station metadata and curated data sets such as earthquake catalogs and phase picks. A database is an organized data collection designed for efficient storage, retrieval, and management. The simplest form of a database, which is termed a flat-file database, relies on files stored in hierarchical directories, where data is typically structured in plain text formats (e.g., CSV, JSON). Although flat-file databases are useful for small-scale or static datasets, their limitations in scalability, query flexibility, and concurrent access become apparent when managing large volumes of data.

Modern databases are broadly categorized into relational and NoSQL (non-relational) databases, each addressing distinct needs. Relational databases were developed decades before the advent of cloud computing. They organize data into structured tables of rows and columns governed by the relational model. The relational model handles structured data with a strict schema and transactional consistency, making it ideal for curated seismic metadata (e.g., event catalogs or station metadata) scenarios.

In contrast, NoSQL databases became popular during the big data era to address the challenges of scalability, schema flexibility, and heterogeneous data types, which are common in modern seismological research. NoSQL databases include document stores (e.g., compatible with a tool such as MongoDB), key-value stores, wide-column stores, and graph databases. The MsPASS framework is an example of using MongoDB to manage large-scale seismic data \citep{wang2022mspass}. NoSQL databases thrive in cloud environments due to their flexibility in managing semi-structured data (e.g., processed waveforms or phase picks) and scalability across distributed systems. This aligns seamlessly with the growing reliance of seismological research on high-volume, multi-modal data sets, such as machine learning-ready archives.

Among NoSQL databases, document stores like MongoDB and its AWS implementation, DocumentDB, exemplify the advantages of schema flexibility in modern seismological workflows. These systems store data as JSON-like documents, enabling researchers to consolidate heterogeneous datasets, such as workflow parameters, phase picks, and semi-structured metadata, into a single database without rigid schema constraints. This flexibility is particularly valuable in seismology, where evolving research workflows often generate metadata with new and inconsistent attributes. DocumentDB further simplifies scalability by automating sharding and replication in cloud environments, allowing distributed storage of large-scale datasets while ensuring low-latency access. Modern cloud databases thus can store the massive output of seismic processing (e.g., billions of phase picks or cross-correlation measurements) and enable quick queries. We will explore the application of cloud-hosted databases in seismology in the following cloud workflow examples.

\section{Cloud Compute}\label{sec:compute}
Cloud providers offer diverse infrastructures and services well-suited to seismologists' diverse needs. Orchestrating these various cloud services to support seismological research involves integrating compute, storage, and software tools provided by cloud vendors using service orchestrators such as CLI (command-line interface) tools, Python-based Software Development Kits (SDKs, e.g., {\tt boto3}, {\tt google-cloud-python}), or web-based platforms (e.g., AWS Step Functions, Google Cloud Workflows). Scientists developing workflows on the cloud face challenges when providers update their services and adapt open-source software, but this is possible with the help of research software engineers \citep{krauss2023seismology}. To detect and adapt to cloud service changes efficiently, ensuring workflow resilience requires version pinning, modular pipeline design, and automated testing.
 
The most basic unit on the cloud for scientists and many other higher-level cloud services is a virtual machine (VM). VMs are the virtualization of hardware and packetization of operating systems, enabling users to access and share physical infrastructure on demand. Different VMs running on the same machine are completely isolated by the host operating system. VMs may be configured at the user's choice, especially to choose the number of vCPUs, RAM, local storage, and additional resources such as GPUs from a set of provisioned templates. The flexibility of VMs is fueling a democratization of large-scale computing. This review focuses on the type of parallelization well suited for cloud platforms, one of distributed memory, sometimes referred to as ``embarrassingly parallel".

\subsection{Batch Computing}
As a service commonly available in Azure, AWS, and GCP, Batch computing is the parallelization of jobs on cloud instances, similar to the job arrays in the SLURM scheduler system \citep{yoo2003slurm}. The Batch computing service is particularly useful for parallelization when the job array shares the same code base and differs only in passing arguments, whether the jobs are simply command lines or containerized tasks. While the embarrassingly parallelizable job runs independently, the multi-node parallel job allows internode communication through message passing libraries \citep[e.g., MPI,][]{gropp1996high}. Such a framework enables single jobs spanning multiple computing instances as a cloud-based on-demand cluster for high-performance computation applications \citep{breuer2019petaflop, zhuang2020enabling, dancheva2024cloud}.

Similar to the scheduler in a modern HPC system, an auto-scaling mechanism dynamically provisions and scales cloud resources based on the volume and requirements of submitted workloads. Such a mechanism adjusts the number of running instances or containers to ensure that resources match the computational requirement without over-provisioning. Auto-scaling can be triggered by pre-defined metrics such as CPU utilization, memory usage, request rates, or job numbers, allowing cloud environments to efficiently handle traffic spikes and workload fluctuations. This capability is essential for maintaining high availability, improving fault tolerance, and optimizing resource utilization and spending, making it a key feature in modern cloud infrastructure.

\subsection{Serverless}
Serverless computing is a cloud-native execution model that abstracts infrastructure management, allowing developers to deploy code without provisioning servers. By automatically scaling resources in response to demand, the serverless architecture enables users to focus on their applications rather than managing operational overhead. This is particularly transformative for seismic early warning systems, where latency-sensitive processing of real-time data (e.g., event detection) requires rapid, event-driven workflows. In a serverless framework, cloud providers like AWS (Lambda), Azure (Functions), and Google Cloud (Cloud Run Functions) dynamically allocate compute resources in response to user requests.

A compelling example of serverless computing in seismology is demonstrated by \citet{mohapatra2025parallel}, who tested a hybrid cloud-local workflow using AWS Lambda and the MsPASS framework. Their study found that downloading raw seismic data (i.e., the 40 million-record USArray dataset) to local HPC clusters created untenable bottlenecks, requiring approximately 462 days for single-worker processing. By shifting preprocessing to serverless functions (e.g., noise reduction, metadata filtering), they minimized data transfer volumes and achieved throughput comparable to local HPC processing. The authors conclude that doing some or all processing on the cloud in this fashion will be essential for any processing involving large volumes of data already stored on the cloud. While hybrid workflows incur cloud costs, they bypass local network limitations, offering a scalable path for modern seismology.

\subsection{Visualization}

Effective visualisation turns today’s petabyte-scale seismic data into insight at a glance. Many groups rely on browser-based notebooks that connect directly to cloud object storage. Browser-based notebooks—commonly hosted in JupyterHub or Google Colab—can mount cloud object storage directly (e.g., using tools such as  {\tt s3fs} for AWS or {\tt gcsfs} for GCP), read only the data chunk required for a plot, and render interactive figures in real time. When those exploratory notebooks mature, researchers often containerise them into lightweight dashboards (such as {\tt Dash} or {\tt Streamlit}) that run on serverless platforms.

Because ‘serverless’ platforms start containers only at researchers' demand (e.g., during an earthquake crisis or a teaching lab), hosting costs remain a few dollars per month in quiet periods. When visualising full 3-D wavefields or large seismic data sets such as from DAS, teams spin up short-lived GPU instances running remote desktop tools such as {\tt ParaViewWeb}, {\tt PyVista}, or cloud-proprietary software such as NICE DCV for AWS, to stream pixels, not raw data, to the user.

Data remains in cloud-optimised formats (Zarr, TileDB) whose data is stored in small, independent blocks (‘chunks’), allowing thousands of viewers to request arbitrary slices without incurring the latency of whole-file downloads. Lightweight assets—thumbnails, map tiles, and prepared iso-surface snapshots—are stored on a content-delivery network for instant access. In contrast, computationally intensive tasks like Fourier transforms, machine-learning phase picking, or on-the-fly iso-surface generation run asynchronously in the background. Those jobs save only small, front-end-ready files (e.g., JSON metadata, PNG images, or glTF models) to object storage, which the web interface retrieves when needed. Seismologists can now interrogate petabyte-scale archives, visualize data by sending only lightweight image tiles to the browser, monitor networks in real time, and teach interactive labs worldwide without local installs or on-prem hardware. Once visual products are in hand, researchers still need robust databases to catalogue the derived picks and images; the next section surveys those options.

\section{Cloud-Native Applications in Seismology}\label{sec:ex}

Seismological analyses often need large computational bursts followed by long idle periods, a usage pattern tailor-made for on-demand cloud resources. Early adopters, therefore, re-hosted existing HPC pipelines in the public cloud to rent capacity only when peaks arose, a form of cloud-assisted ``lift-and-shift" workflow. \citet{wang2018public} was among the first to demonstrate the use of public cloud computing for seismic data processing at the TB scale and performed noise cross-correlation using the Aliyun cloud service, specifically with the Batch service and the cloud object storage. \citet{maccarthy2020seismology} used the FDSN web service to request data on the fly while detecting harmonic tonal noise. They successfully scanned 6 TB of USArray data within 4 days on a Kubernetes cluster with 50 EC2 nodes. \citet{witte2020event} built a serverless seismic imaging application ported from the HPC platform, dynamically scheduling jobs and provisioning computational resources using serverless cloud,  demonstrating excellent cost efficiency, scalability, and performance competitive with on-premise HPC clusters. Similarly, \citet{zhu2023quakeflow} designed an integrated earthquake detection workflow with containerized submodules, i.e., data streaming, phase picking, association, and event location. The auto-scaling mechanism was implemented at both the cluster and cloud platform levels, enabling the automatic provisioning of computational resources based on job load. The paradigm of these cloud-based workflows was summarized by \citet{maccarthy2020seismology} as comprising three primary components: the infrastructure, cluster management software, and domain-specific research software. These ``lift-and-shift" studies share a three-tier pattern: infrastructure (cloud VMs and containers), cluster management layer (Kubernetes, batch, step functions), and domain software (e.g., noise cross-correlation, full waveform inversion, earthquake detection). Computing efficiencies depends on how tightly these elements are coupled.

\begin{figure}
    \centering
    \includegraphics[width=\linewidth]{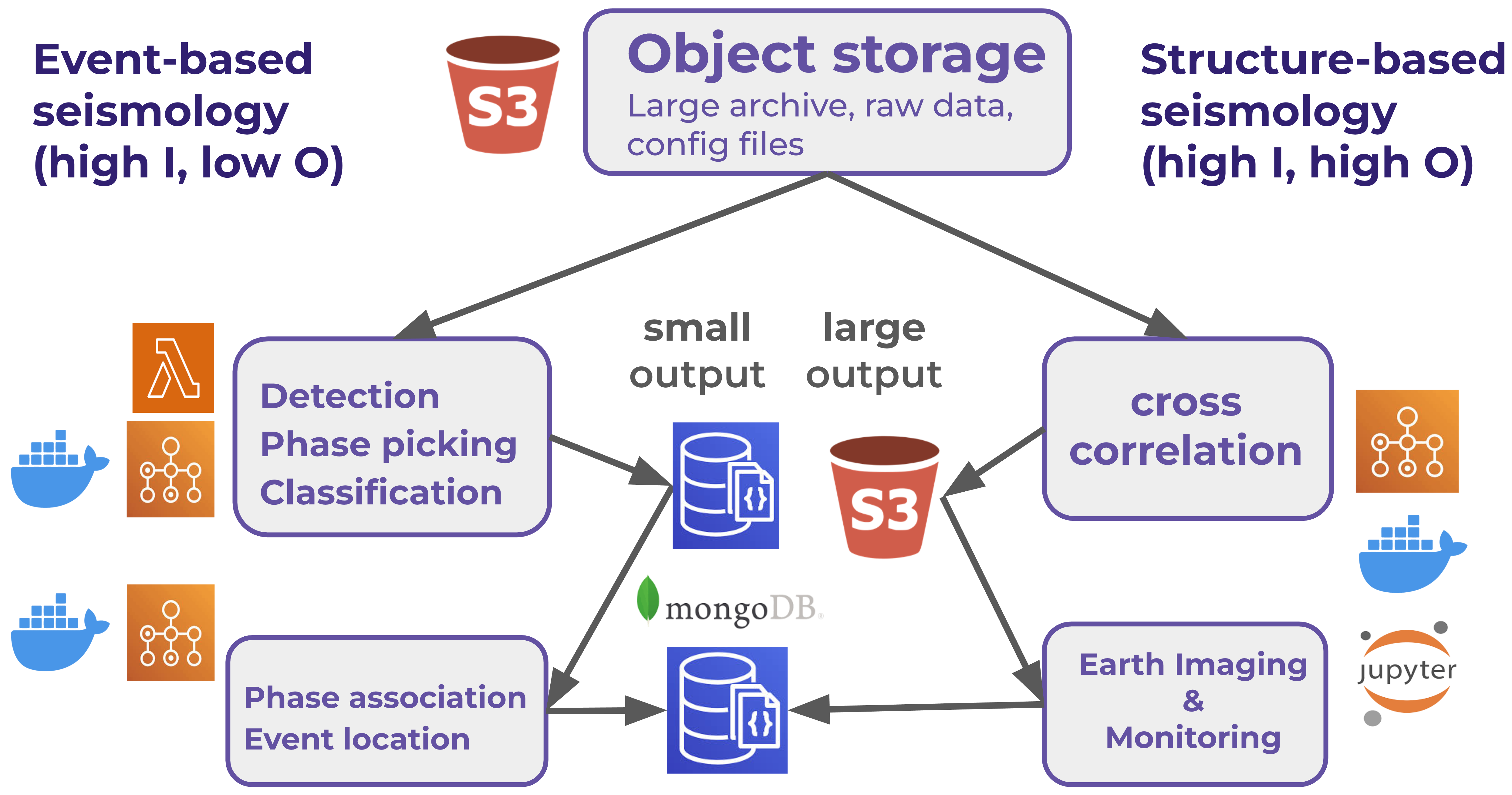}
    \caption{{\bf Two canonical workflows in seismology:} Event-based seismology that reads large volumes of seismic data (high I - high Input) but outputs low volumes of data in databases (low O), and structure-based seismology that reads large volumes of seismic data (high I) and outputs large volumes of seismic data (high O). The first workflow outlines the basic steps in generating a seismic event catalog. The second workflow describes how to extract seismic properties of the subsurface with ambient field seismology, which generates high data volumes of ambient noise cross-correlations.}
    \label{fig:workflows}
\end{figure}

However, these workflows assumed that data could be fetched on demand, but pulled data from outside the cloud; such a prerequisite cannot be easily met in the cloud for data-intensive tasks. Transferring the raw data or data products can be time-consuming and expensive \citep{wang2018public, ni2023object}. For example, \citet{zhu2023quakeflow} spent $\sim$50\% of total job time downloading waveforms through the FDSN web service. Requesting data at scale may also pose challenges for data centers that receive unpredictable heavy traffic and clusters where big data is saved and managed on-site. To overcome these bottlenecks, recent effort embrace ``cloud-native" workflows: 1) harness direct access to the cloud-hosted data to avoid copying data \citep{maccarthy2019putting, yu2021scedc}, 2) leverage cloud managed services, and 3) employ containerized software to ensure portability, consistency, and ease of deployment across platforms. To illustrate the cloud-native workflows for seismology, we present two contrasting research workflows in seismology that have benefited from the cloud systems: 1) cross-correlation for ambient noise seismology and 2) earthquake catalog building workflows. Figure~\ref{fig:workflows} illustrates the two alternative workflows with data flows and associated cloud services best used in large-scale jobs.

\subsection{Workflow 1: Large Ambient Noise Seismology}\label{sec:exp1}
Ambient noise seismology is the methodology that utilizes continuous seismograms, typically dominated by a diffuse, ambient seismic field, to extract spatial or temporal variations in seismic wave speeds. The main advantages of the method have been to 1) Earth Imaging reconstruct high-frequency Rayleigh wave and image shear wave structure where seismic stations are located \citep[e.g.][]{shapiro2005resolution} and no longer rely on rare earthquakes, and to 2) Earth monitoring by exploring changes in subsurface structure by subtle phase shifts on the coda of cross-correlations \citep[e.g.,][]{Sens-Schönfelder2011passive}. 

The method relies on the cross-correlation of short time series between channels, presenting some of the most significant challenges in computational seismology. The cross-correlation typically uses short windows, ranging from minutes to a few hours, recorded at pairs of seismic channels, and stacks these over days to years of data. Thus, the workflow scales quadratically with the number of channels {\it N}, a step that favors shared memory processing, and linearly with the number of windows to stack {\it T}, a step that favors distributed memory processing. The rise of array seismology with $N > 100$ is a real computational challenge. Because the workflow often entails storing cross-correlation functions, including intermediate steps such as substacking, cross-correlation may involve writing TBs of files. Several efforts have been made for open-source and large-scale computing of ambient noise cross-correlations, some leveraging CPU-based clusters \citep[e.g.,][]{jiang2020noisepy, makus2024seismic}, others leveraging heterogeneous computing with CPU and GPU \citep[e.g.,][]{fichtner2017supercomputer, clements2020seisnoise, ventosa2019towards, zhou2021high}. 

Given two canonical seismological approaches for {\it Earth imaging} and {\it Earth monitoring}, the workflow to compute cross-correlation functions is multi-step, and their optimal parallelism strategies differ. First, the cross-correlations are performed independently on synchronous time series, which permits distributed memory parallelism, often referred to as "embarrassingly parallel," and scales only with the overall period of the instrumental record. Second, the cross-correlations are done on {\it pairs} of seismic channels, and a given channel window of data could be read once and cross-correlated over all other channels with $N(N-1)/2$ pairs. This strategy often employs multi-threading with shared memory, parallelization across channel pairs, and leveraging GPUs to accelerate the correlation step \citep{fichtner2017supercomputer, clements2020seisnoise}. When the data is too large for the available memory, local storage of intermediate products, such as the Fourier transforms \citep{wang2018public}, or low-rank factorizations \citep{martin2019scalable}, and parallelization over groups of station pairs is also possible \citep[e.g., C4 project][]{schmitt2025c4, schmitt2020ground}.

Cloud infrastructure is particularly well-suited for ambient noise seismology, given its significant data throughput (reading and writing) and parallelization capabilities. Several attempts to perform ambient noise cross-correlations on the cloud demonstrated the speed and scalability of adapting cloud infrastructure. \citet{wang2018public} developed a parallelization scheme to independently calculate groups of channel pairs and perform massive daily cross-correlations, totaling 300M, over 10 hours of processing on nearly 1000 virtual machines using the Aliyun cloud service. \citet{ni2023object} performed DAS cross-correlations on AWS using cloud-native workflows and achieved 300M of daily cross-correlations over 64 instances in 24 hours, spending less than \$20. \citet{clements2020scedc} and \citet{schmitt2020ground} developed a workflow on AWS that approached "cloud-native" by streaming from S3 to EC2, generating cross-correlation locally, saving the results on disk (using EBS storage), and uploading them to S3.

We now present {\bf Cloud-Native NoisePy}, a new version of \citet{jiang2020noisepy} that has been updated with I/O for cloud-based data archives, enhanced object-oriented Python programming, including parallelization, flexibility in computing platforms, and continuous integration. NoisePy leverages S3 for massive I/O parallelization and short-term storage of temporary data, such as daily cross-correlation, as illustrated on the right-hand side of Figure~\ref{fig:workflows}. NoisePy employs two primary parallelization strategies to optimize performance. The first leverages the AWS Batch services with FARGATE serverless compute to execute concurrently and independently each daily job of processing and inter-channel cross-correlation using the same container but different data to ingest. Within each job, NoisePy utilizes native Python multi-threading for parallelization across several steps, including reading data, pre-processing, computing the Fourier transform, cross-correlation, and writing daily results, which are stacked and saved as compressed NumPy {\tt .npz} files back to S3. After processing all daily data, a final aggregation step combines the results to produce long-term correlation stacks. This architecture efficiently handles large-scale data, generating TBs of cross-correlation outputs in the cloud.

\begin{figure}
    \centering
    \includegraphics[width=\linewidth]{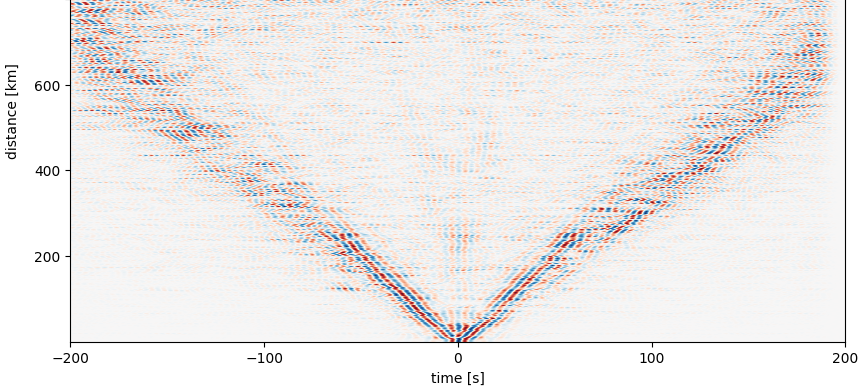}
    \caption{{\bf Ambient Noise Cross Correlation} using 1 year of data in the Southern California Seismic Network: all data are publicly available in the scedc-pds S3 bucket. The data is filtered between 1 and 10 seconds.}
    \label{fig:noisepy}
\end{figure}

We present the results of an experiment in which we ran NoisePy on one year of SCEDC data. We cross-correlated all of HH? channels that included 288 stations and about 43K station pairs. Ran on Fargate with up to 64 instances, it took 11 hours of compute time to generate 6.2M files of daily-stacked cross-correlation, with a volume of 1.6 TB on S3, spending about \$250 on SPOT pricing. The second step on a similar Fargate cluster took 1.5 hours and generated the final stacks of all inter-channel cross-correlations, totaling 23 GB of data and 46,000 S3 objects. We present the results of the 1-year stacked ZZ component of the cross-correlation, filtered 5-10 seconds, sorted by the inter-station distance in Figure~\ref{fig:noisepy}. We find the convergence of the correlation functions past 600 km of inter-station distances. Seismic tomography from these data products will involve extracting phase and group velocity measurements from these cross-correlations and inverting the frequency-dependent velocity curves into a shear-wave velocity model. Our experiment demonstrates the case of scalable data processing for tomography applications.  

\subsection{Workflow 2: Earthquake Catalog Building}\label{sec:exp2}

Earthquake catalog building is a complicated, multi-step workflow that ingests raw time series data and outputs point clouds of earthquake locations and their attributes. The main steps are to detect events, identify the time at which seismic phases (typically {\it picking} P and S arrivals) arrive, associate them with a specific origin (event), possibly incorporate location using 3D Earth velocity models, relocate them using double-difference relocation, and calculate source parameters, such as magnitude and focal mechanisms. Each step has been explored using machine learning. In particular, for phase picking, deep learning has proven highly successful, with models such as U-Net \citep[e.g.,][]{zhu2019phasenet, munchmeyer2022which}. Network-based analysis typically requires gathering multiple-station data simultaneously, and has benefited greatly from U-Net and graph networks \citep[e.g.,][]{munchmeyer2021transformer, sun2023phaseno, clements2024grapes}. Most workflows are a sequence of modules \citep[e.g.,][]{zhang2022loc, walter2020easyquake, retailleau2022wrapper, zhu2023quakeflow}, where modules can be adapted according to the user preferences.

The computational efficiency of these workflows matters when considering large-scale deployment. Similar to the first workflow presented, we break down the computational efforts into two types of parallelization. The first is the extraction of features from raw data, such as the arrival times of P and S, the wave amplitude, and perhaps the polarity of the P and S waves, a process that can be independently calculated on each window of data as an independent job and can be parallelized massively. The second processing set requires aggregating these features across the stations and benefits from multi-threaded parallelization. Both steps can be easily orchestrated on cloud systems, which was first pioneered by \citet{zhu2023quakeflow} by including deep learning phase picking, association, and relocation using double difference, and by \citet{pierleoni2023iot} for Internet-of-Things early warning systems whereby picking is done at the seismometer level, and location is done on the cloud. We present here another cloud-native workflow for the basic steps of earthquake catalog building, utilizing the {\tt SeisBench} ecosystem \citep{woollam2022seisbench} for phase picking and event discrimination, which we refer to as QuakeScope.

QuakeScope orchestrates its seismic catalog workflow entirely on AWS Batch (Fargate) and a MongoDB-compatible DocumentDB cluster for outputting detection attributes and checkpointing. The basic unit of the workflow is a Python job, composed of four steps: (i) obtain a day-long time series of miniSEED data, (ii) process the continous waveform with a phase picking model implemented in SeisBench, such as \citep[PhaseNet][]{zhu2019phasenet} or \citep[EqTransformer][]{mousavi2020earthquake}, and a classifier that predicts four classes (earthquake, explosions, surface events, and noise) as a form of detection (Kharita et al, in prep), (iii) remove the instrumental response to extract the amplitude of each detection, (iv) write the resulting detections to a DocumentDB. 

As jobs (i) and (iv) are I/O-bound, while jobs (ii) and (iii) are compute-bound, we implemented an asynchronous processing using the asyncio Python module. This means that the job processes data simultaneously, loads the next day, and still writes the picks from the previous day. This increases throughput substantially and thereby reduces resource costs. The individual steps communicate through limited-size First-In-First-Out queues to avoid excessive memory overheads.

\begin{figure}
    \centering
    \includegraphics[width=1\linewidth]{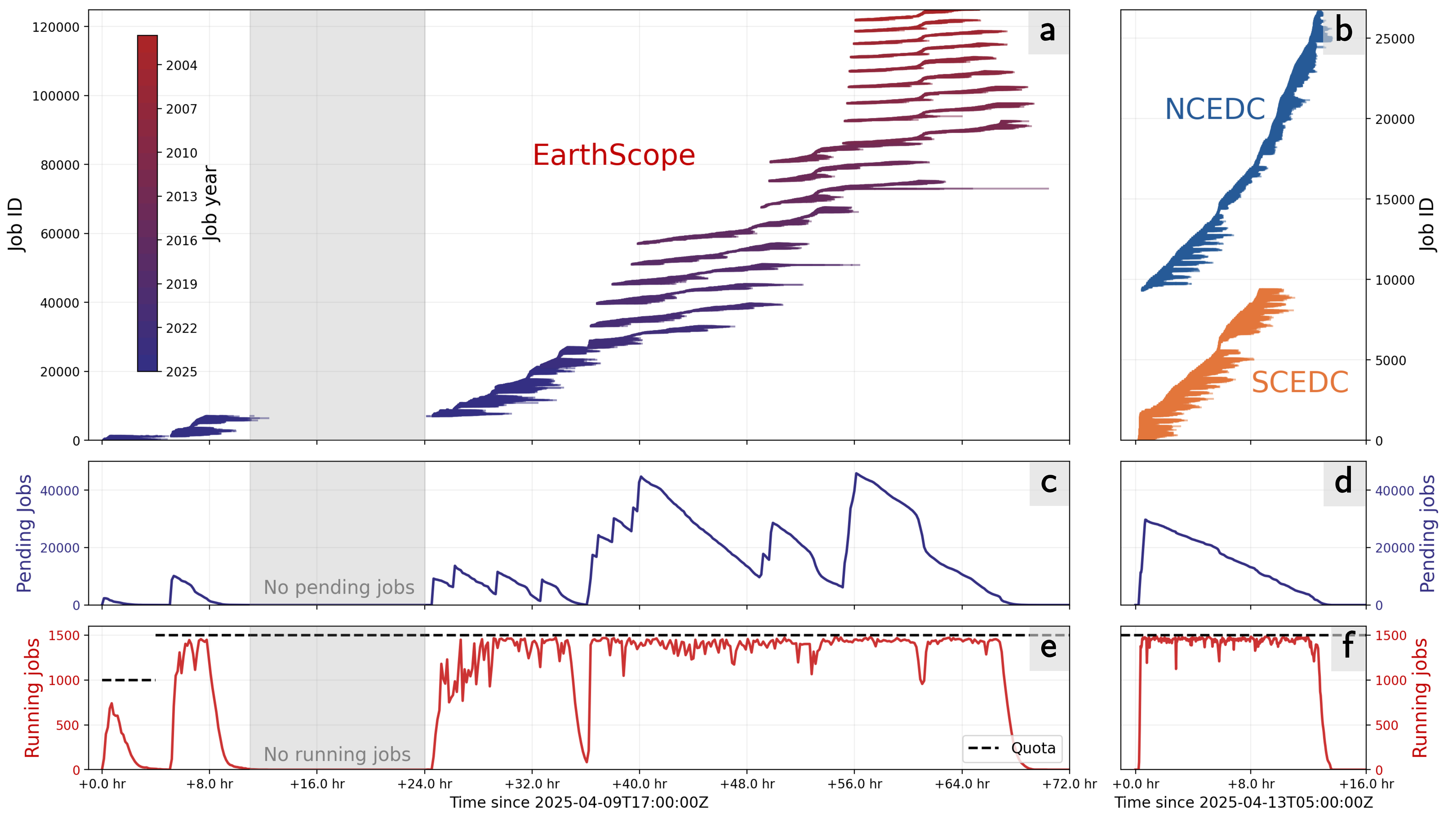}
    \caption{{\bf Evolution of job statistics in the 2025 Data Mine experiment}: top panel has a line between the start and end time of each job that ran on Fargate. Each year's worth of data is sent manually as a separate batch of jobs, color-coded for EarthScope. The middle panel shows the number of pending jobs in the Fargate queue. Jobs enter the queue and are scheduled to run until the quota of 1500 jobs is reached. The dips in the number of active jobs are attributed to our slower job orchestration. One that may be attributed to SPOT's intermitency is at +60Hr. Upon success, the experiment on the NCEDC and SCEDC went smoothly, maximizing the number of jobs until completion.}
    \label{fig:datamine2025}
\end{figure}

We present the results of a 2025 Data Mine experiment. We scanned the complete EarthScope Consortium-managed miniSEED archive (1~PB) and the SCEDC and NCEDC AWS-hosted open data, each with 150 TBs, with QuakeScope. We describe in Figure~\ref{fig:datamine2025} the evolution of our jobs on AWS. Our AWS allocation had limited quotas for 1,500 jobs (12,000 vCPUs) to be used simultaneously, a tripled quota relative to the default limits. We manually launch each year, from 2025 going back in time, as a queue of jobs, and experimented with the EarthScope back-end servers as a pilot experiment. During the initial stage, we stress-tested the EarthScope archive.  We progressively launched the jobs on Fargate Spot (a lower-cost queue that is less "on-demand"). In the EarthScope experiment, our progressive load of jobs in the queue demonstrated the resilience of the backend system, allowing us to launch a larger batch of jobs by hour 40 in the experiment. For the NCEDC and SCEDC experiments with a smaller dataset of approximately 300 TB, we launched all jobs simultaneously, reached our quotas, and completed them in 12 hours. The performance of cloud systems remained constant, demonstrating that this process can be easily accelerated by allowing for more concurrent processes. 

\section{Community Model Visualization and Dissemination}\label{sec:viz}

Within the seismological community, there is an acute need for models of natural systems to support both basic and applied research. For example, models of seismic wave velocities within the Earth are essential for interpreting tectonic history and evolution \citep{delph2021subcretionary}, for generating accurate ground motion estimates using earthquake simulations \citep[e.g.,][]{glehman2024partial}, and for enabling a wide range of analyses related to seismic hazard and risk. Similarly, geometric models of earthquake-producing fault structures, along with associated metadata such as slip rates and earthquake histories \citep{seebeck2024new, plesch2024scec}, form the foundation of seismic hazard analysis \citep{petersen2020update}. These needs have driven the development of community models, which are collaboratively developed, synthesis-based representations of natural systems that are maintained and shared by a broad group of researchers with expertise in the relevant domain \citep[e.g.,][]{aagaard20252024californiacommunityearth}.  Community models are typically open-source and publicly available, designed to incorporate the best available data and understanding, and intended to serve as common foundations for scientific research, education, and practical applications \citep[e.g.,][]{shaw2015unified, plesch2024scec}.  The generation of community models is a focus of various community science organizations such as the Statewide California Earthquake Center (SCEC) and the Cascadia Region Earthquake Science Center (CRESCENT) \citep{melgar2024cascadia, aagaard20252024californiacommunityearth}.

To enable FAIR (Findable, Accessible, Interoperable, and Reusable) access to community models \citep{wilkinson2016fair} and facilitate their use across a wide range of applications, these models are often distributed in multiple formats—such as structured text files, binary volumes, databases, or web-based APIs—to accommodate diverse user needs and computing environments. However, the complexity and size of these datasets—particularly 3D and 4D models of Earth's interior or fault systems—can present significant technical barriers for many users \citep{small2017scec}. There is a growing need for software tools that allow users to interactively visualize, query, and subset models before downloading or integrating them into workflows. These tools must balance performance with accessibility and support interoperability with widely used scientific programming languages (e.g., Python) and data standards. Ultimately, the effective use of community models depends not only on their scientific rigor but also on the availability of user-friendly software that lowers the barrier to entry.

The cloud is particularly well suited for the hosting, visualizing, and disseminating community models because it provides centralized, flexible infrastructure that meets the needs of distributed scientific teams \citep{gentemann2021science}. Community models are often large, dynamic data, and are accessed by users across multiple institutions. These characteristics make local storage solutions inefficient or inaccessible. Cloud storage enables elastic scaling, allowing both storage and computing power to dynamically adapt to the community model's growth, without the managerial burdens associated with on-premises infrastructure. It also facilitates consistent versioning, access control, and metadata management, which are required for transparency and reproducibility. In addition, cloud platforms integrate seamlessly with modern computational tools and workflows, enabling users to analyze and visualize models directly in the cloud without transferring large datasets. As such, the cloud is a natural fit for managing community-driven, data-intensive geoscientific resources.

As an example of such a tool in the seismological community, the CRESCENT Community Velocity Model (CVM) Viewer and Repository \citep{bahavar2025cvm} is a cloud-based platform for storing, distributing, analyzing, and visualizing seismic wave velocity models of the Earth. It combines Python-based tools with a geospatial web interface to enable real-time, interactive exploration of datasets. By adhering to widely accepted metadata and file format standards, the platform ensures that datasets remain consistent, interoperable, and ready for use in both research and education.

The CVM Viewer is a web-based geospatial visualization tool built with Python, FastAPI, and CesiumJS \citep{cesium2018cesium}. It allows users to explore CVM datasets interactively through a cloud-hosted 3D map. In addition to visualizing the spatial extent of seismic velocity models, the viewer displays known faults and earthquake hypocenters. Users can toggle terrain layers, adjust model boundaries, and navigate using rotation, zoom, and pan controls. Visualization tools include horizontal slices, vertical cross-sections, and depth profiles, offering intuitive ways to investigate subsurface structures.

The CVM Repository hosts multiple user-submitted seismic velocity models that have undergone peer review and are published. Models are stored in netCDF-4 Classic and HDF5 formats on AWS S3, organized hierarchically to separate 3D model volumes and associated surface data. Automated compliance checks are performed before storage to ensure alignment with community metadata standards. The backend, built with xarray and h5py, handles data queries and retrieves model subsets based on user-defined geographic and depth ranges.

Users can extract horizontal slices, cross-sections, or full 3D volumes in various supported formats. These extraction and conversion tools are deployed using AWS Lambda, enabling efficient access to large netCDF or h5 files and seamless format conversion. Throughout these operations, geospatial metadata is preserved to ensure compatibility and compliance with metadata standards. The entire system is deployed using AWS Fargate, a serverless container platform that automatically scales computing resources in response to user demand. 

In addition to the CVM Viewer, CRESCENT is developing a suite of complementary cloud-based tools to support the broader earthquake science community. The CRESCENT Community Fault Model Viewer \cite{bahavar2025cfm} enables interactive visualization and dissemination of fault geometries and associated metadata. Other tools currently under development include platforms for storing, analyzing, and distributing paleoseismic data and seismicity catalogs, all designed with an emphasis on scalability, accessibility, and adherence to community data standards.  These emerging tools demonstrate how cloud infrastructure makes data and computational resources more accessible and promotes collaboration across disciplines and institutions.

\section{Education and Training Initiatives}\label{sec:edu}
Open-source software and interactive cloud-based computing environments have transformed seismological research and education by providing free, accessible data analysis and modeling tools. Platforms like Binder, Google Colab, and institutional JupyterHub instances enable researchers and students to run open-source Jupyter Notebooks without requiring local installation, significantly lowering entry barriers. {\tt Seismo-live} \citep{krischer2018seismo} is one example, offering a library of seismology-focused Jupyter Notebooks that can be executed directly in a web browser using Binder. Similarly, Google Colab provides a free cloud-based notebook environment with pre-installed libraries, enabling students to analyze seismic datasets and run numerical simulations from any device. Google Colab is often linked in seismological software repositories to provide tutorials on free cloud services \citep[e.g., in SeisBench,][]{woollam2022seisbench}. By leveraging these free Jupyter-based platforms, the seismology community ensures that computational tools are widely available, fostering open science, reproducibility, and equitable access to high-performance research workflows. 

Research projects may also lead to developing cloud-based workflows, and researchers may choose to provide guides for cloud usage. For example, \citet{krauss2023seismology} utilized the Azure platform and compared pre-trained Machine Learning models and template matching for constructing an earthquake catalog offshore. Their work also provided informative, educational guidance for individual researchers in code development and containerization, cloud infrastructure, job design, and performance analysis.

In higher education, instructors leverage cloud platforms to create interactive, scalable learning environments for seismology and data science. Universities may deploy cloud-like infrastructure with virtual machines for classroom instruction, enabling students to access the course with affordable devices (e.g., tablets, laptops, or even phones) and run code and homework on cloud instances. Such cloud integrations enhance accessibility, enabling students to work with real-world data and modeling problems in an educational setting.

Hands-on training workshops are essential for advancing computational seismology skills, as they immerse participants in using modern software and large-scale computing resources. Recent community initiatives, such as the NSF-funded SCOPED project, have organized multimodal workshops to teach researchers and students how to use research-grade seismological software on cloud platforms \citep{denolle2024training}. Engaging the community in multiple ways with cloud infrastructure has been beneficial: from running simple workflows on a provided cloud-based JupyterHub (e.g., the GeoLab workshop) to deploying an EC2 instance on their own (e.g., in \href{https://seisscoped.org/HPS-book/intro.html}{HPS}). Recent workshops have been dedicated to training several hundred participants, primarily graduate students, postdoctoral researchers, and research scientists.

\section{Discussion and Outlook}\label{sec:discussion}
\subsection{Software as a service}
Cloud infrastructure lets seismologists rent powerful computing services only when needed, but those virtual servers still start life as “bare-bones” operating systems. Researchers must rebuild their software stack--libraries, compilers, scripts--every time they launch a new instance. To simplify this process, the common practice is to replicate working environments facilitated by environment management software, such as a lightweight solution of the package manager Anaconda (\url{https://anaconda.org}), or utilize a fully self-contained option such as Docker \citep{merkel2014docker} and Singularity \citep{kurtzer2017singularity}. Cloud platforms also provide ready-to-use virtual images that cater to general or geophysics-specific needs, often at a small additional cost.

A more user-friendly approach is Software-as-a-service (SaaS). It is a delivery model that enables users to run scientific software easily in the cloud while interacting through a web form or API. The backend software is cloud-optimized, and computing resources are provisioned upon users' request. Such a model enables users to access the software in a serverless setting without tedious configuration, while being elastic and cost-efficient for service providers. For example, \citet{chen2013cloud} proposed a web application that allows users to submit requests to generate synthetic seismograms. The service receives requests along with source parameter settings and initiates the 3D elastic wave equation solver on the backend. Researchers of interest may utilize this service through a direct and convenient web interface, receiving synthetic seismograms without requiring any software configuration. The trade-off is flexibility—custom methods or novel algorithms still require direct access to code and data, which SaaS platforms may not expose.

\subsection{The self-imposed research reproducibility}
Seismology already enjoys a culture of open data and open-source software—community archives expose waveforms through standard FDSN web services, and libraries such as ObsPy \citep{beyreuther2010obspy} and SeisBench \citep{woollam2022seisbench} make analysis scripts widely shareable. Yet, the full research replication requires running the full-stack workflow on any machine to obtain similar research results.

Cloud computing has become a critical enabler of reproducible research by forcing workflows to be explicit and portable. In traditional observational seismology, workflows often rely on trial and error, such as manually selecting frequency band filters based on domain expertise and visual data inspection. While this approach benefits from expert judgment, it poses significant challenges to reproducibility — other researchers may struggle to replicate results if they do not use the same parameters or follow the same steps. Cloud computing, however, requires the creation of well-defined, containerized workflows that are portable and executable in consistent environments. This shift toward standardized, containerized research workflows facilitates reproducibility by ensuring that the exact conditions under which the research was conducted can be easily replicated on different systems or by different researchers. This self-imposed reproducibility fosters a more rigorous and transparent scientific process, crucial in modern seismology, where big-data applications increasingly dominate.

\subsection{Cost of the cloud}
Most cloud service providers employ a pay-as-you-go pricing model, where users are only charged for using any related cloud resource. Researchers can estimate the cost of their workload to the first order by timing the duration of time that virtual machines run and the time that data is stored in cloud storage. For example, we summarize the spending of \citet{clements2023signature} based on AWS EC2 and S3 pricing policy: 1) downloading and uploading 50 TB of NCEDC data ($\sim$\$40), 2) performing the single-station noise correlations ($<$\$50), and 3) storing all data over one week ($\sim$\$40/week). Such pricing model holds cost advantages: 1) cloud resources are accessible to almost everyone, whilst on-premise equipment makes one-time spending unnecessary, 2) maintenance is performed by cloud service providers instead of full-time institutional IT employees, and 3) spending is better quantified and monitored through the billing statistics and may help future budgeting \citep{norman2021cloudbank}. 

However, the pricing model can be complex when chargeable usages are vaguely defined for some services and resources. Consequently, budgeting for cloud infrastructure beyond the basics is often more complicated. For example, in the cloud-native NoisePy test (see Section \ref{sec:exp1}), we utilize AWS S3 to save pre-stacked correlation functions before stacking. Besides the ephemeral storage cost from pre-stacked correlation functions, S3 write (PUT) and read (GET) operations also come at a flat rate (usually several USDs per thousand requests). Despite minimal cost, such cost shall not be omitted when performing a large-N and large-T ambient noise interferometry study with potentially millions of similar S3 requests. Moreover, data bytes traveling across cloud regions will incur a perceivable egress cost, unless waived under specific agreements (e.g., the Open Data Sponsorship Program, which supports the SCEDC S3 archive presented by \citet{yu2021scedc}). Additional charges may also be applicable for on-demand servers. For instance, I/O operations with the standard pricing model are not free for DocumentDB clusters. With this type of cluster, spending may be significant when I/O usage is extensive (e.g., phase picks, insertions, and metadata queries that scale with the job, see Section \ref{sec:exp2}). Such scaling terms should be identified and optimized to avoid unexpected spending in a cloud-native workflow. Despite these challenges, our experience has been fortunate to overbudget, strategize to optimize, and conclude with a lower overall cost.

\section{Conclusion}
The central message of this review is simple: thanks to cloud object storage, elastic compute, and containerized workflows, analysing petabyte-scale seismic archives is no longer aspirational—it is routine. Tasks that once clogged campus networks and monopolised HPC queues now complete in hours or days, with costs that fit a student budget. Building a cloud-ready pipeline remains a significant undertaking for most scientists. Containerizing legacy code, wiring up object storage I/O, and automating provenance all require skills that fall between classical research and production software. Therefore, collaboration with research software engineers remains essential: they translate scientific intent into robust, version-controlled artifacts, add automated tests, and keep pace with the rapid evolution of cloud services. Once that up-front investment is made, every run is self-documenting and trivially repeatable, lowering the long-term maintenance burden. Prototyping for cloud systems on-premise is beneficial and possible, for example, using tools such as MinIO object storage systems and MongoDB databases. Researchers may focus on workflow reproducibility, open-sourcing software, and minimizing job requirements to increase speed and lower computing and environmental costs.

Our performance benchmarks illustrate the payoff. In our ambient-noise test case, 1.6 TB of correlations—6.2 million files—were generated in eleven hours for about \$250, all without touching a single on-premise disk. Likewise, the QuakeScope catalog builder scanned roughly a petabyte of global data in three days, limited only by computing quotas, and likely discovered 10 times more earthquakes than previously reported. Automatic retries on pre-emptible (“Spot”) instances kept utilisation high and human intervention low, demonstrating that today’s managed services can rival dedicated HPC for embarrassingly parallel seismology at comparable cost.

Looking ahead, both DAS and the rapid proliferation of low-cost IoT seismometers will significantly increase data volumes and latency requirements, far exceeding today’s norms. In an Earthquake Early Warning architecture, hundreds to thousands of edge devices can run lightweight pickers locally, then stream only compact, parametric data to the cloud. There, serverless functions fan-in those messages, trigger association and localization, and broadcast alerts—often within a few seconds of the origin time. The rapid scaling of cloud resources is particularly suitable for handling rare and extreme events. Finally, Cloud providers’ per-region carbon-intensity dashboards (e.g., AWS’s Sustainability Pillar, Google’s Carbon-Free-Energy scores, Microsoft’s 2025 renewables target) can inform users to choose cloud computing regions with a lower carbon footprint without sacrificing latency and performance.

Realizing that vision will take community effort—shared, containerized code, FAIR data in cloud-optimised formats, and collaborations between domain scientists and research software engineers. But the heavy lift is now within reach. The petabyte era is not a looming burden. Embracing cloud and other scalable computing solutions transforms this burden into a catalyst. It allows researchers to interrogate Earth processes at resolutions and timescales that were previously out of reach, opening new frontiers of discovery. It also permits researchers to focus on fundamental physical processes, rather than being limited by a given observational period and spatial extent. Petabyte-scale seismology is now practical.

\begin{acknowledgments}
This work is supported by the Seismic Computational Platform for Empowering Discovery (SCOPED) project under the National Science Foundation (award numbers OAC-2103701 (UW), OAC-2103494 (UT)). The Schmidt Futures Foundation also supported the development of NoisePy at the University of Washington's Scientific Software Engineering Center. The EarthScope Consortium, through a Pass-Through Entity (PTE) Federal award no 2310069, partially supported this work. EarthScope data were accessed from the NSF SAGE data archive operated by EarthScope Consortium (award number 1724509). The computing resources presented in this paper were obtained using CloudBank \citep{norman2021cloudbank}, which is supported by the National Science Foundation (award number CNS-1925001). The Harvard Data Science Initiative supported the development of the Julia Cloud workflow, developed by T. Clements and J. Schmitt, NSF EAR-1850015 award. JM has been funded by the European Union under the grant agreement n°101104996 (“DECODE”). We are grateful for discussions with Manochehr Bahavar and Loïc Bachelot surrounding visualization on the cloud.
\end{acknowledgments}

\begin{dataavailability}
The SCEDC and NCEDC data used in this study are publicly available through the AWS Open Data Sponsorship Program (\url{https://registry.opendata.aws/southern-california-earthquakes/} for SCEDC and \url{https://registry.opendata.aws/northern-california-earthquakes/} for NCEDC). The EarthScope Consortium data could be accessed through the EarthScope FDSN web service. Software in Section~\ref{sec:ex} are available at \url{https://github.com/noisepy} and \url{https://github.com/seisSCOPED/quakeScope/}.
\end{dataavailability}

\appendix

\label{lastpage}

\bibliographystyle{gji}

\end{document}